# Odd spin symmetry and anisotropy switching in *p*-wave magnet CeNiAsO

Fayuan Zhang[1,2*#], Huaxun Li[3*], Xingkai Cheng[4*], Yibo Fan[1*], Yifan Yin[1], Yifan Gao[4], Zhanfeng Liu[5], Shengtao Cui[5], Zhouyi Yin[6], Yue Zhao[6], Junhao Lin[6], Zhengtai Liu[7], Mao Ye[7], Yaobo Huang[7], Shan Qiao[8], Wu Xie[9], Ping Miao[9], Hao Wu[1#], Junwei Liu[4#], Guanghan Cao[3#] & Chaoyu Chen[1,6#]

[1] Songshan Lake Materials Laboratory, Dongguan, China.

[2] Quantum Science Center of Guangdong-Hong Kong-Macao Greater Bay Area, Shenzhen, China.

[3] School of Physics, Zhejiang University, Hangzhou, China.

[4] Department of Physics, The Hong Kong University of Science and Technology, Hong Kong, China.

[5] National Synchrotron Radiation Laboratory, University of Science and Technology of China, Hefei, China.

[6] Department of Physics, Southern University of Science and Technology, Shenzhen, China.

[7] Shanghai Synchrotron Radiation Facility, Shanghai Advanced Research Institute, Chinese Academy of Sciences, Shanghai, China.

[8] Shanghai Institute of Microsystem and Information Technology, Chinese Academy of Sciences, Shanghai, China.

[9] Spallation Neutron Source Science Center, Dongguan, China.

[*] These authors contributed equally to this work.

[#]Correspondence should be addressed to F.Z. (zhangfayuan@quantumsc.cn), H.W. (wuhao1@sslab.org.cn), J.L. (liuj@ust.hk), G.C. (ghcao@zju.edu.cn) and C.C. (chenchaoyu@sslab.org.cn).

**Abstract:**

**Odd-parity magnets, complementary to altermagnets, exhibit unique properties such as high efficiency in charge-spin conversion and compatibility with conventional superconductivity, of critical importance in the pursuit of energy-efficient spintronics and topological superconductors for quantum computation. For even-parity *d*-wave and *g*-wave altermagnets, the magnetic structure, spin-split band structure and physical properties are currently under intensive study. On the contrary, while dozens of odd-parity magnets and various promising properties have been predicted in theory, experimental studies are scarce. Specifically, the magnetic structure and transport properties of candidates $NiI_2$ and doped $Ga_3Ru_4Al_{12}$ have been reported, yet the characteristic band structure and particularly the odd-parity spin symmetry remain elusive. Here we demonstrate experimentally the deterministic *p*-wave spin symmetry and resistance anisotropy switching for the prototype odd-parity magnet, CeNiAsO. Angle-resolved photoemission spectroscopy (ARPES) reveals two cleaved terminations with distinct surface band structure. By compensating the polar surface, we achieve intrinsic bulk band structure, for which the spin splitting can be well described by the *p*-wave magnetic structure through first-principles calculation. The bulk spin polarization measured by spin-resolved ARPES exhibits symmetry with only one degenerate plane, fingerprint of *p*-wave magnetism. We further report giant resistance anisotropy and demonstrate switching between high-resistance and low-resistance states through modest field-induced domain selection, highlighting its potential for antiferromagnetic spin memory devices. The structural similarity between CeNiAsO and 1111-type Fe-based superconductors stimulates further exploration on the interplay between *p*-wave magnetism, superconductivity and band topology.**

# I. Introduction

Spin-split magnets with compensated moments not only promise scalable, energy efficient, ultrafast functionality for information storage and logic, but also breeds exotic quantum phenomena bridged to superconductivity, topological phases and ferroelectricity, greatly enriching the research landscape of fundamental magnetism and spintronics[1-6]. Described by distinct spin groups, these unconventional magnets are characterized by compensated magnetic moments in lattice space and non-relativistic, spin-split band structure in reciprocal space[7-19]. According to the symmetry of spin-momentum locking, these magnets are primarily categorized into even-parity[11,13,19,20] and odd-parity classes[21-26]. For even-parity magnets such as $d$-wave altermagnets $RbV_2Te_2O$ family[11,27-31] and $g$-wave altermagnet $\alpha$-MnTe[13,32-36] family, their magnetic structure, spin-split band structure and physical properties are currently under intensive study[37-40], composing the most active field in condensed matter physics.

Odd-parity magnets[21-26], complementary to altermagnets, exhibit unique properties such as high efficiency in charge-spin conversion[23] and compatibility with conventional superconductivity[26,41], of critical importance in the pursuit of energy-efficient spintronics and topological superconductors for quantum computation. $P$-wave spin symmetry, as the simplest odd-parity spin splitting, was originally discussed in the form of spin-channel Pomeranchuk instability[7,8,42]. Recent advances proposed alternative route to realized $p$-wave spin splitting promoted by suitable non-relativistic lattice and spin symmetries, termed $p$-wave magnetism[21]. In the prototype magnetic structure (Fig. 1a, Model 1), the non-collinear coplanar spin arrangement is described by a non-relativistic spin symmetry $[C_{2\perp}\|\vec{t}]$, where $C_{2\perp}$ is a 180° spin-space rotation around the axis perpendicular to the coplanar plane and $\vec{t}$ is a lattice-space translation. Combined with the noncentrosymmetric magnetic structure, such symmetry allows for compensated magnetic order with zero magnetization and $p$-wave spin-momentum locking in the band structure (Fig. 1b).

Theoretical study of odd-parity magnets is starting to thrive with dozens of materials candidates predicted [19,21,24], yet experimental advances remain few. Unlike the simplest $2 \times 1$ magnetic supercell (Fig. 1a, Model 1) in the pioneering proposal [21], recent experimental efforts focus on spin helices (Fig. 1a, Model 2) with large and complicated magnetic cell doubling, commensurate[43] or incommensurate[44]. While the magnetic structure and anisotropic transport properties (schematically shown in Fig. 1c) have been experimentally explored[43], the characteristic odd spin symmetry of the band structure bridging the former two characteristics remain elusive, not limited to $p$-wave helix magnet, but for all the odd-parity magnets.

Here we demonstrate the deterministic odd spin symmetry and anisotropic transport behavior in the first theoretically proposed *p*-wave magnet CeNiAsO. Unifying bulk band structures are achieved for both CeO- and NiAs-terminations, through compensating the polar surface states via *in situ* potassium deposition in spatial-resolved ARPES measurement. Combining ARPES, spin-resolved ARPES and first-principles calculation, we directly demonstrate the anisotropic band spin-splitting and particularly the spin polarization symmetric with respect to only one nodal plane, characteristic of *p*-wave order. Together with the tetragonal symmetry-breaking resistance and anisotropy switching, these findings establish a prototype *p*-wave magnet with promising spintronic applications.

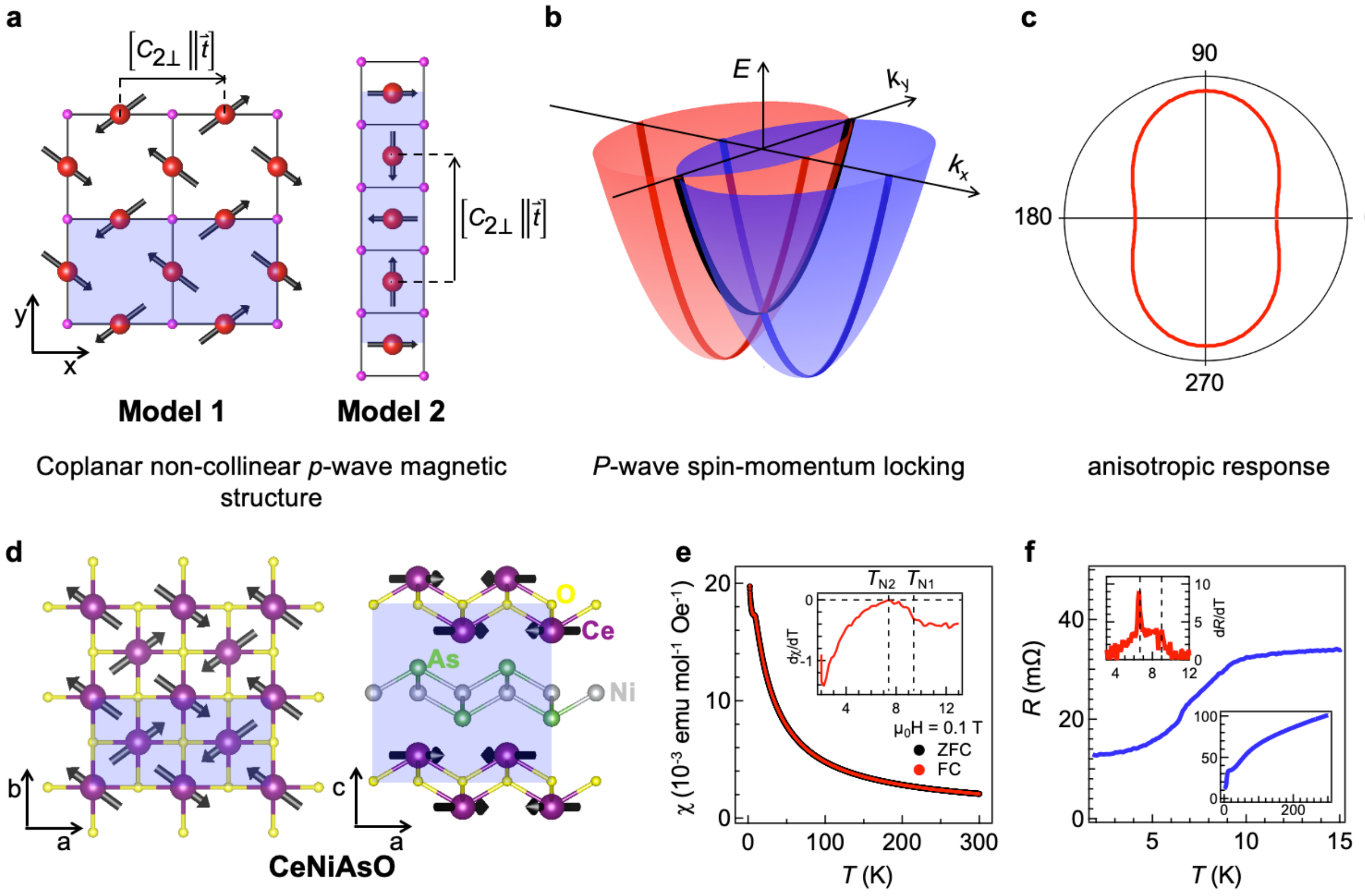

**Fig. 1 | Characteristics of *p*-wave magnet and properties of CeNiAsO. a-c,** Schematic illustration of the *p*-wave magnetic structure models (**a**), *p*-wave spin-momentum locking (**b**) and anisotropic response (**c**). **d**, Magnetic structure of CeNiAsO at $T < T_{N2}$. **e**, Zero-field-cooled (ZFC) and field-cooled (FC) magnetic susceptibility (χ) for CeNiAsO polycrystals with 0.1 T field. Inset: the derivative of χ(*T*) highlights two transitions at $T_{N1} = 9.4\ K$ and $T_{N2} = 7.4\ K$. **f**, Temperature dependent resistance in the range of $2\ K - 15\ K$ (main) and $2\ K - 300\ K$ (right bottom inset). Left top inset: first derivative of *R*(*T*). The transition temperature shows slight sample difference.

## II. Magnetic structure, termination dependent surface states, intrinsic bulk bands

CeNiAsO adopts the ZrCuSiAs structure (space group $P4/nmm$). Single crystals in millimeter size have been grown (Methods), and the layered, tetragonal crystal structure is confirmed (Extended Data Table 1 and Fig. 1). As shown in Fig. 1d, the lattice consists of conducting NiAs layer and insulating CeO layer stacking along the *c*-axis. This compound was recognized as an antiferromagnetic (AFM) Kondo metal, with quantum critical point achieved by pressure or doping [45,46]. At low temperature, two AFM transitions, $T_{N1}{\sim}9.4\ K$ and $T_{N2}{\sim}7.4\ K$, can be resolved from the temperature dependent magnetic susceptibility (Fig. 1e) and resistance curves (Fig. 1f). Systematic magnetic structure studies[47,48] suggest an incommensurate spin density wave state for $T_{N2} < T < T_{N1}$ and a coplanar $2 \times 1$ AFM order at $T < T_{N2}$. The magnetic moments are primarily from Ce site without contribution from Ni atoms. Shown in Fig. 1d, the ground state AFM order doubles the tetragonal crystallographic unit cell along the $a$-axis (blue shaded region), with the noncollinear Ce moments lying in the $ab$-plane. According to the theoretical proposals[21,23], such magnetic order is described by nonrelativistic spin group symmetry $[C_{2\perp}||\vec{t}]$ (Fig. 1a, Model 1) and allows for out-of-plane spin polarization with single nodal plane in the Brillouin zone (BZ) (Fig. 1b). Such $p$-wave spin symmetry leads to spontaneous anisotropy of the transport responses (Fig. 1c), as electrical resistance is lower for current flowing parallel to the nodal plane than that perpendicular to it. While the coplanar magnetic structure[48] and anisotropic resistance[49] have been observed in CeNiAsO, the deterministic spin symmetry of $p$-wave magnetism awaits verification.

The layered stacking structure of CeNiAsO naturally leads to two cleaved terminations, i.e., the positively charged CeO-termination and negatively charged NiAs-termination. Using spatial-resolved ARPES with $\sim 10\ \mu m$ beam spot size, the termination dependent core level spectra (Extended Data Fig. 2) and low-energy electronic structure (Figure 2) have been directly measured. Both terminations host Fermi surface with similar features, including sharp, circular pockets surrounding M points and faint, elliptical features at X/Y points (Fig. 2a-d). Photon energy-dependent spectral mapping show negligible dispersion along $k_z$ direction (Extended Data Fig. 3), in line with its layered structure. According to our first-principles calculations (Extended Data Fig. 4) based on the $p$-wave order, the low-energy bulk bands consist of two intertwining band pairs $\alpha$ and β. The nonsymmorphic symmetry enforces the double degeneracy of these bands along the BZ edge M-X (without spin-orbit coupling). Except that, they split into $\alpha_{1/2}$ and $\beta_{1/2}$ pairs. On the Fermi surface, bulk $\alpha$ and β bands form the two circular pockets surrounding M, while the split $\alpha_1$ band and β band form the ellipses at X/Y.

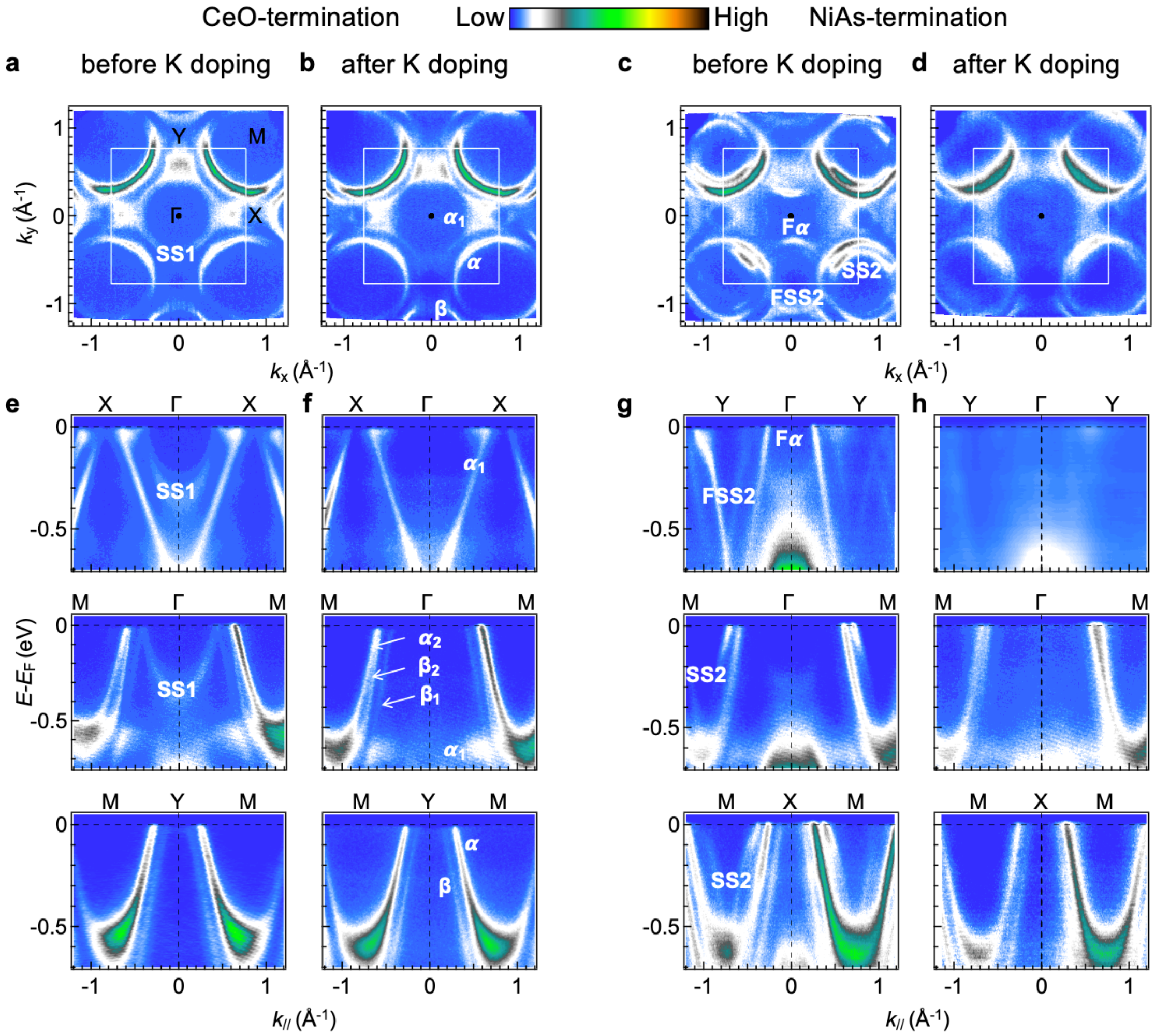

**Fig. 2 | Termination dependent surface state bands and the intrinsic bulk bands of CeNiAsO. a-d,** ARPES measured Fermi surfaces (6.5 K) for as-cleaved (**a**), K-doped (**b**) CeO-termination and as-cleaved (**c**), K-doped (**d**) NiAs-termination, respectively. **e-h**, Corresponding band spectra along high-symmetry paths. Abbreviations: $\alpha$/ $\alpha_1$/$\alpha_2$/β/$\beta_1$/$\beta_2$: bulk bands; SS1/SS2: surface states; FSS2: folded surface state 2; F$\alpha$: folded bulk $\alpha$ band.

Besides the bulk $\alpha$ and β bands, the polar surface after cleavage induces additional surface states, similar to the case in isostructural LaFeAsO [50]. For CeO-termination, we observed one additional surface state SS1, which makes up a parabolic electron pocket centered at Γ (Fig. 2a and 2e). For NiAs-termination, another surface state SS2 appears at M (Fig. 2g), forming a circular Fermi pocket with strongly anisotropic intensity distribution inside the $\alpha$ pocket (Fig. 2c). Furthermore, a $2 \times 1$ lattice reconstruction occurs at this termination (see Extended Data Fig. 5 for microscopic data), causing band folding from M to Y point, as evidenced by both the folded surface state FSS2 and the folded bulk $\alpha$ band F$\alpha$.

To eliminate the complexity brought by the surface polarity/reconstruction and get access to the intrinsic, $p$-wave bulk bands, we developed a potassium (K) deposition approach to compensate the charge gap between the cleaved surface and bulk. Such approach turns out to be effective for both terminations. As shown in Fig. 2b, 2f for K-doped CeO-termination and Fig. 2d, 2h for K-doped NiAs-termination, the SS1 and SS2 features have been cleared off, such that both terminations present intrinsic bulk Fermi surfaces and band spectra without qualitative difference. Furthermore, the effect of $2 \times 1$ lattice reconstruction at the NiAs-termination has also been effectively suppressed, judging from the barely visible folded bands at the Y point (Fig. 2h). Such clarification is indispensable as the targeted $p$-wave magnetism forms the same $2 \times 1$ lattice doubling, so that the structural origin should be excluded before talking about the spin symmetry.

## III. Spin splitting and spin polarization of *p*-wave symmetry

Intrinsic spin splitting and polarization demonstrating the $p$-wave symmetry remain the missing link connecting $p$-wave magnetism and anisotropic physical responses in the surging study of odd-parity magnets. Here we tackle this by jointing ARPES, spin-resolved ARPES and density functional theory (DFT) analyses. The ground $p$-wave magnetic state of CeNiAsO has a $2 \times 1$ unit cell (Fig. 1d), breaking the tetragonal symmetry in the paramagnetic phase. The corresponding calculated Fermi surface presents clear odd parity spin symmetry, i.e., $s_z(k_x) = -s_z(-k_x)$ (Fig. 3d). Yet the corresponding ARPES Fermi surface (Fig. 3b, $T = 6.5\,K$ below $T_{N2}$, after K doping) is still dominated by tetragonal symmetry as in the paramagnetic phase. This suggests weak band folding by the magnetic unit cell doubling, similar to its superconducting counterpart CeFeAsO [51]. Consequently, the calculated Fermi surface is unfolded to the $1 \times 1$ BZ as shown in Fig. 3e, restoring the tetragonal band symmetry while still preserving the $p$-wave spin symmetry. Comparing Fig. 3b and 3e, one immediately realizes that all the bands forming Fermi surfaces are spin polarized except at the $k_x = 0$ nodal plane. Clear spin splitting can be resolved, especially between the $\alpha$ and β bands surrounding M as long as $\alpha_1$ around X.

To directly visualize the $p$-wave spin polarization symmetry, we present the measured spin-resolved energy-distribution curves (EDCs) and the corresponding spin polarization component $s_z$ in Fig. 3g, for 4 momenta $k_{1,2,3,4}$. Focused on the spectra (band $\alpha_1$) along $X' - \Gamma - X$ (Fig. 3c, right panel), these momenta cross the $\alpha_1$ band at binding energy ~0.35 eV, manifested as broad peaks in the spin-resolved EDCs (highlighted by dashed boxes). For all the 4 momenta measure at $T < T_{N2}$, slight asymmetry between red and blue channels can be resolved, corresponding to weak spin polarization $s_z$ indicated by the red and blue filled regions. As expected, the measured $s_z$ is red (negative) for negative momenta $k_{1,2}$ and is blue (positive) for positive momenta $k_{3,4}$, respecting the *p*-wave symmetry. On raising the sample temperature above $14\ K$, while the band spectra remain unchanged due to the local *c*-*f* type interaction (Extended Data Fig. 6), the asymmetry in spin-resolved EDC disappears (Extended Data Fig. 7), suggesting zero spin polarization at the paramagnetic phase and the correspondence between the *p*-wave spin symmetry and magnetism. The comprehensive agreement between (spin-resolved) ARPES measurements and DFT calculations provides direct spectroscopic evidence of *p*-wave spin symmetry.

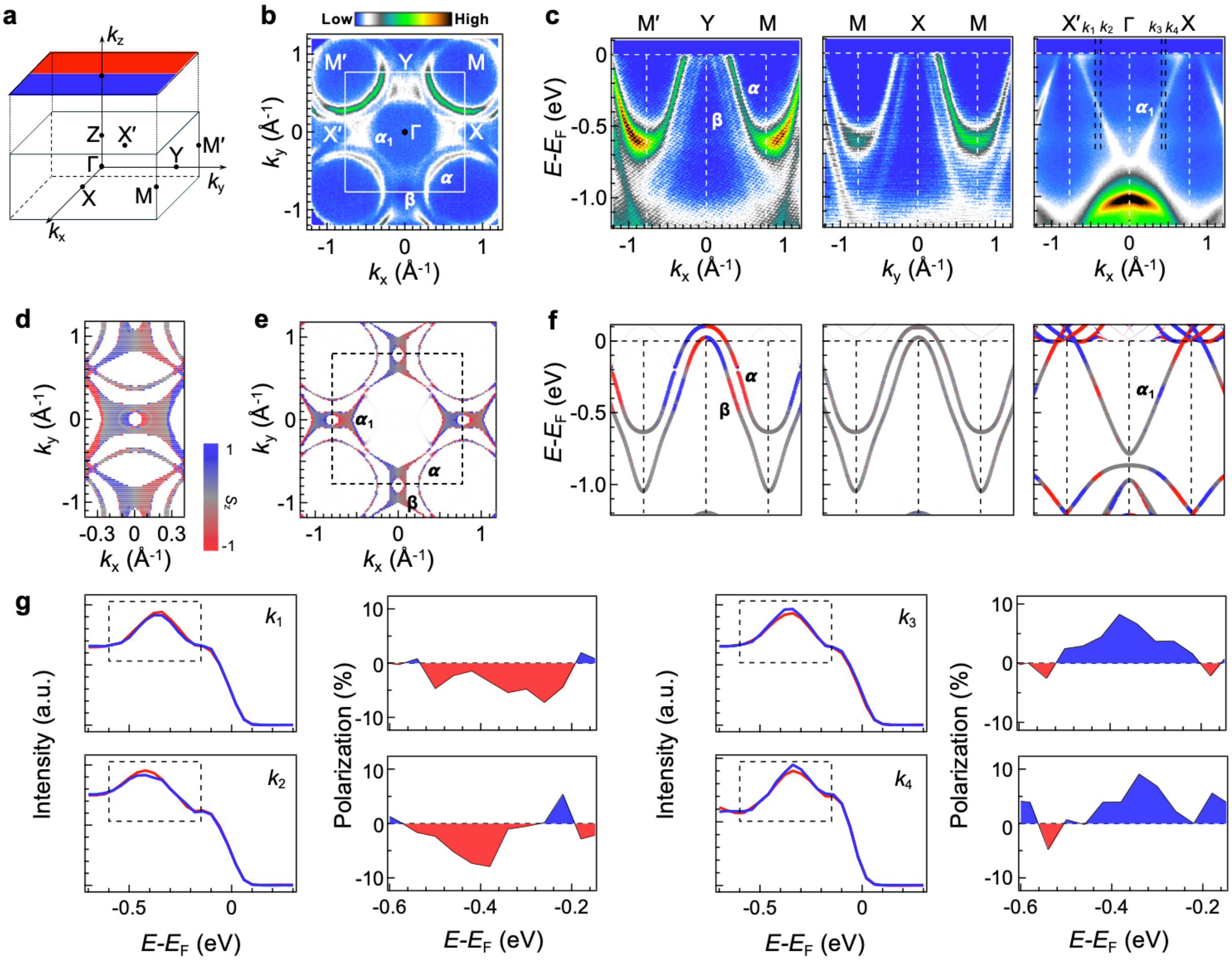

**Fig. 3 | *P*-wave spin symmetry in CeNiAsO. a,** Bulk BZ and the projected surface BZ with p-wave symmetry. **b-c**, Fermi surface and band spectra along high-symmetry path, measured at T = 6.5 K below $T_{N2}$. **d**, Calculated spin polarized Fermi surface based on the $2 \times 1$ *p*-wave magnetic order. **e**, Same as **d** but unfolded to the $1 \times 1$ BZ. **f**, Calculated spin-polarized dispersion corresponding to the measured ones. **g**, Spin-resolved ARPES results (T = 6.6 K below $T_{N2}$) for the selected momenta $k_{1,2,3,4}$ as indicated in the $\mathrm{X}' - \Gamma - \mathrm{X}$ spectra in **c**. Red and blue curves are the measured spin-resolved EDCs, while the color-filled curves are the corresponding spin polarization.

## IV. Anisotropic resistance, field-selected domain population and resistance switch

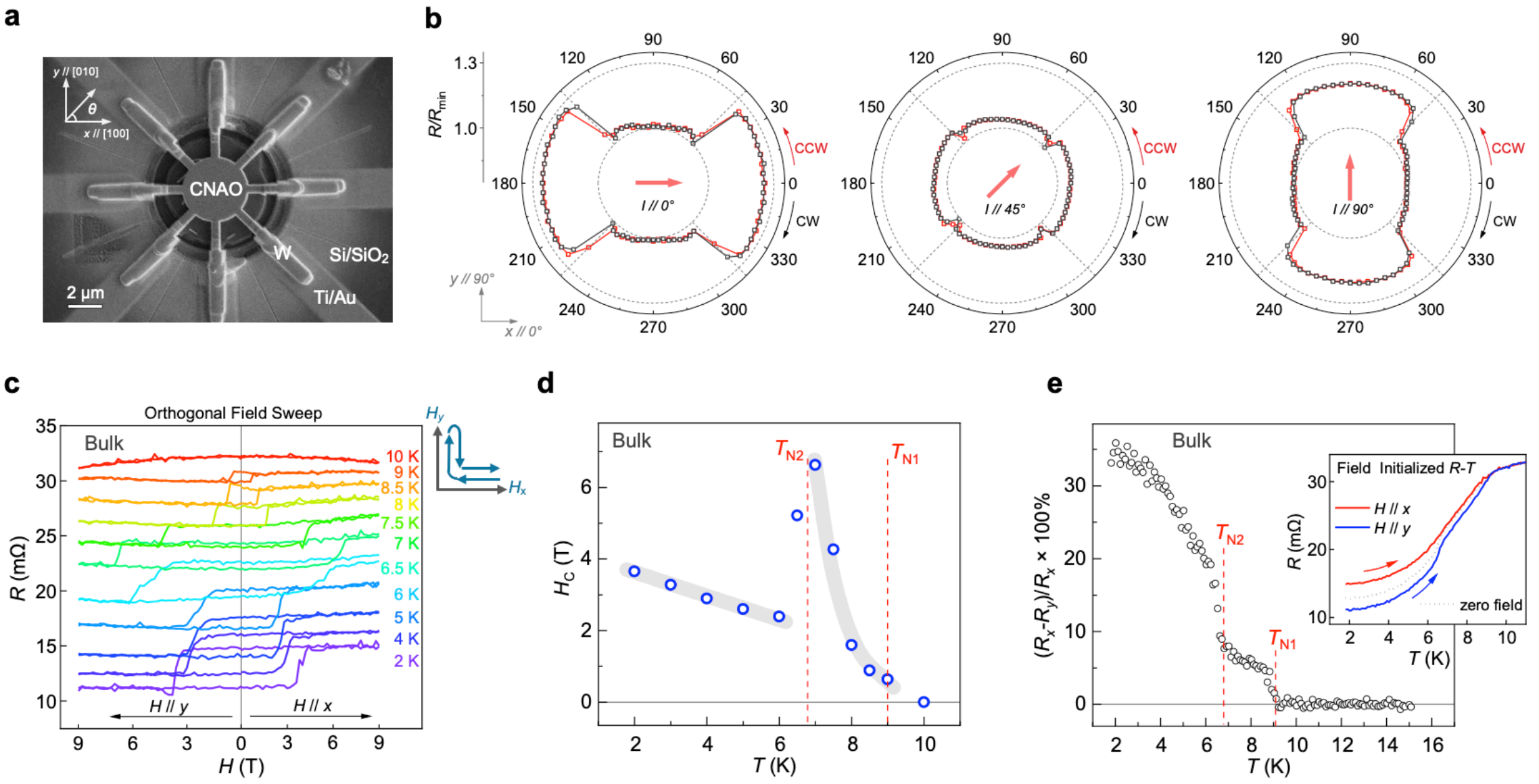

**Fig. 4 | Nonvolatile, anisotropic resistance of CeNiAsO. a.** Scanning electron micrograph of the FIB fabricated Device A. The $x$-axis is defined as $x \parallel a$. During transport measurements, the current is applied along 0°, 45° or 90°, and an in-plane magnetic field $H$ is applied at an angle $\theta$, with respect to $x$-axis. **b**. Polar plot of the $\theta$-dependent $R/R_{min}$ under a 12 T in-plane magnetic field at 4 K. The red (black) curve represents the magnetic field sweeping counterclockwise (clockwise). **c**. Magnetoresistance hysteresis loops of the bulk device measured at different temperatures. The arrows in the coordinate system indicate the scanning direction of the orthogonal magnetic field, namely along $x$-axis and $y$-axis. **d**. The temperature-dependent coercive field ($H_C$) extracted from **c**. The gray lines indicate that the temperature dependence of $H_C$ can be divided into two distinct regimes. **e**. The inset shows the $R$-$T$ curves of bulk CeNiAsO single crystal measured along $x$-axis after a 6 T initializing magnetic field applied along $x$-axis (red) or $y$-axis (blue), and the arrows represent the temperature sweeping direction. An initialized field is applied at 2 K and removed during the measurements. The dash line is the $R$-$T$ curve measured without applying initialized field. The temperature-dependent magnetoresistance ratio calculated by $(R_x\text{-}R_y)/R_y\times 100\%$ is shown in the main panel, where the $R_x$ ($R_y$) represent the resistances measured after an initializing magnetic field applied along $x$-axis ($y$-axis).

To explore the potential spintronic functionality of $p$-wave magnets, we investigated the anisotropy of the resistance in focused-ion-beam (FIB)-fabricated CeNiAsO devices (Device A in Fig. 4a and Device B in Extended Data Fig. 8) and bulk samples (bulk device in Extended Data Fig. 8). The resistance is systematically studied by varying the angle $\theta$ between the in-plane magnetic field and the current direction during measurements. As shown in Fig.4b, as the magnetic field angle changes, the resistance switches between two stable states and shows a two-fold symmetry when the measurement currents are applied at 0° and 90°. The $\theta$-dependent anisotropic resistances exhibit a hysteresis behavior between clockwise (CW) and counterclockwise (CCW) magnetic field scanning, which is a typical feature of magnetic orders. Significantly, the angular dependence of the magnetoresistance measured under these two current configurations exhibits a 90° phase shift relative to each other. This demonstrates two key points. First, under magnetic field rotation, one of the two degenerate antiferromagnetic domains ($2 \times 1$ and $1 \times 2$) with mutually orthogonal sublattice orientations is selectively stabilized, resulting into two distinct resistive states for a given current direction. Second, for a given domain population, the electrical scattering condition is highly different for these two measurement currents: a high-scattering channel for current along one direction and a low-scattering channel for current along the perpendicular direction. As schematically show in Fig. 1b-c, low-resistance state is expected for current flowing parallel to the nodal plane (parallel spin scattering) while high-resistance state corresponds to current perpendicular to the nodal plane (antiparallel spin scattering). For current applied along 45°, the band splitting along the $x$ and $y$ directions are equivalent; therefore, no pronounced anisotropic resistance is observed.

Further temperature-dependent magnetoresistance measurements show that the hysteresis behavior persists below $T_{N1}$= 9 K in both bulk device (Fig. 4c) and FIB device B (Extended Data Fig. 8), which is consistent with the emergence of magnetic order below $T_{N1}$. The magnetoresistance ratio can reach as high as 35% in bulk device and 48% in FIB device. Due to the $p$-wave symmetry, a 180° switching of domain population generates the same state of magnetoresistance, we then conduct a 90° magnetic field sweep to visualize such process. As shown in Fig. 4c, when the magnetic field is applied alternatingly along $x$-axis and $y$-axis, a clear hysteresis loop of the magnetoresistance is observed, which resembles the hysteresis loop characteristic of ferromagnetism. Remarkably, the high remanence ratio of the bulk device demonstrates a robust resistance state in CeNiAsO. The temperature-dependent coercive field ($H_C$) of the bulk device is extracted from the hysteresis loops and summarized in Fig. 4d. The hysteresis loop behavior persists up to $T_{N1}$, and the evolution of $H_C$ can be divided into two temperature regimes by $T_{N2}$. Below $T_{N2}$, $H_C$ increases approximately linearly with decreasing temperature, whereas above $T_{N2}$, the temperature dependence becomes nonlinear. The magnetoresistance ratio (Fig. 4e) can also be separate into two distinct regimes by $T_{N2}$. Based on the above results, an intimate connection between the *p*-wave magnetic order, *p*-wave spin symmetry and the large nonvolatile anisotropic resistance has been established, demonstrating its great potential for applications in antiferromagnetic memory devices.

## V. Discussion

Through comprehensive structural, band structure, spin polarization and resistance studies employing the multifold functionalities of ARPES and transport techniques, we have achieved access to the intrinsic bulk band spin splitting and spin polarization respecting the *p*-wave symmetry in CeNiAsO, completing the essential missing link connecting *p*-wave magnetism and anisotropic physical responses in the surging study of odd-parity magnets. The significancy of demonstrating *p*-wave spin symmetry in CeNiAsO is multidimensional. While altermagnets preclude $Z_2$ topological insulator due to time-reversal symmetry breaking, odd-parity magnets including *p*-wave ones allow for $Z_2$ classification as the existence of effective time-reversal symmetry. Although topological surface band structure awaits ARPES detection, signature of Dirac fermions has been revealed in FeAs-based isostructural antiferromagnets [52]. Importantly, pressure and substitutions on the rare earth, 3*d* metal or O sites have been tested effective to modulate the magnetic structure[53], trigger the heavy-fermion antiferromagnetic quantum criticality[46] and even introduce superconductivity with the highest critical temperature in Fe-based superconductors [54]. Thus, there are multiple approved approaches available to explore the entanglement of unconventional magnetism, band topology and superconductivity in this material family and our observation of *p*-wave spin symmetry in CeNiAsO would certainly accelerate such adventure.

## Methods

### Sample growth and characterization

Single crystals of CeNiAsO were grown using a NaAs-flux technique. Polycrystalline CeNiAsO was first synthesized by a solid-state reaction method[46]. NaAs flux was prepared by reacting Na and As chunks at 600 °C. The CeNiAsO precursor and NaAs flux were mixed in a molar ratio of 1:5, loaded into an alumina crucible, sealed in a Ta tube, and subsequently encapsulated in an evacuated quartz ampoule. The assembly was heated to 1150 °C over 1000 min in a muffle furnace, held at this temperature for 24 h, and then cooled to 600 °C at a rate of 2 °C $h^{-1}$. After furnace cooling to room temperature, residual flux was removed by rinsing the crystals with deionized water.

X-ray diffraction measurements were carried out at room temperature on a PANalytical diffractometer (Empyrean Series 2) with Cu-$K_{\alpha 1}$ radiation, and a Bruker D8 Venture diffractometer with Mo $K_\alpha$ radiation. The chemical composition of single crystals was determined by energy-dispersive X-ray spectroscopy (EDS, Oxford Instruments X-Max) equipped in a scanning electron microscope (SEM, HITACHI S-3400N). Magnetic properties were measured using a Quantum Design Magnetic Property Measurement System (MPMS3).

### ARPES and spin-resolved ARPES measurements

ARPES measurements were performed at the BL03U beamline of the Shanghai Synchrotron Radiation Facility (SSRF) equipped with a Scienta Omicron DA30 energy analyzer and *p*-polarized radiation. During the experiment, the beam spot size was set to $15\times15$ $\mu m^2$. The energy resolution was set to 10-20 meV depending on the photon energy used, and the angular resolution was set to 0.2°. The spin-resolved ARPES experiments were performed at National Synchrotron Radiation Laboratory (NSRL) BL13U. Spin-polarizations of photoelectrons were detected by two very-low-energy electron diffraction (VLEED) spin detectors which can detect two spin components, where one component is in-plane and parallel to the ARPES slit and the other is out-of-plane. Complementary ARPES measurements were performed at the BL09U of SSRF.

### First-principles calculations

The calculations were performed in the framework of density functional theory as implemented in Vienna ab initio simulation package[55]. The projector-augmented wave potential was adopted with the plane-wave energy cutoff set at 460 eV. The exchange-correlation functional of the Perdew–Burke–Ernzerhof type has been used for both structural relaxations and self-consistent electronic calculations, with the convergence criteria as $10^{-5}$ eV[55,56]. The Brillouin zone was sampled by a $4 \times 8 \times 4$ Γ-centered Monkhorst–Pack mesh. The spin–orbit coupling strength was recompiled and set to zero to simulate the nonrelativistic p-wave spin splitting. The magnetic moments were fixed along the (±0.3, ±0.22, 0) direction to match the μSR experimental data. We performed a band-unfolding procedure[57] to obtain the effective band structure and Fermi surface for direct comparison with the ARPES experiments.

**Device fabrication and measurements**

The focused-ion-beam (FIB) devices were fabricated by a Helios 5UX FIB system. The Ti/Au (Ti 5 nm/Au 15 nm) electrodes were pre-patterned on Si/$SiO_2$ substrates using a standard photolithography process. A lamella was extracted from a bulk CeNiAsO flake and thin down to approximately 500 nm. Then, the lamella was transferred onto the substrate, and FIB-induced W deposition was employed to establish electrical connection between the sample and Ti/Au electrodes. For the bulk device, CeNiAsO thin flake and hBN are transferred using PDMS stamps onto the Ti/Au electrodes. The transport measurements are performed in a physical-property measurement system (PPMS, Quantum Design) using the built-in transport measurement option.

**ACKNOWLEDGEMENTS**

C.C. is supported by the National Key R&D Program of China (grant nos. 2025YFA1411200 and 2022YFA1403700), the National Natural Science Foundation of China (NSFC) (grant no. 12574068), Guangdong Basic and Applied Basic Research Foundation (grant nos. 2026B0303000004, 2022B1515020046, 2022B1515130005 and 2021B1515130007).

**Author contributions**

C.C. conceived the idea and proposed the research. F.Z., Z.F.L, S.C., Z.T.L., M.Y., Y.H. and S.Q. contributed to the (spin-)ARPES instruments, measurement and analysis. H.L., Z.Y., Y.Z., J.H.L., W.X., P.M. and G.C. contributed to the sample growth and characterization. X.C., Y.G. and J.W.L performed calculations/simulations. Y.F., Y.Y. and H.W. performed the resistance measurements and analysis. All authors wrote and corrected the manuscript.

## Competing interests

The authors declare no competing interests.

**Correspondence and requests for materials** should be addressed to Chaoyu Chen.

## Code availability

The codes for analyzing the data of this study are available from the corresponding authors upon request.

**Extended Data Table 1 | Crystallographic data of CeNiAsO obtained through structure refinement of single crystal x-ray diffraction at 300 K (final *R* indexes [I ≥ 2σ(I)] *R*1 = 0.0115, *wR*2 = 0.0280; final *R* indexes [all data] *R*1 = 0.0115, *wR*2 = 0.0280).**

| Chemical formula | | | | | CeNiAsO | |
|---|---|---|---|---|---|---|
| Space group | | | | | P4/nmm (No.129) | |
| a (Å) | | | | | 4.0773 (2) | |
| c (Å) | | | | | 8.1198 (7) | |
| atoms | Site | x | y | z | U | Occ. |
| Ce | 2c | 0.5 | 0 | 0.35376 (4) | 0.0063 (2) | 1 |
| As | 2c | 0 | 0.5 | 0.14373 (7) | 0.0113 (2) | 1 |
| Ni | 2b | 0.5 | 0.5 | 0 | 0.0127 (2) | 1 |
| O | 2a | 0 | 0 | 0.5 | 0.0072 (8) | 1 |

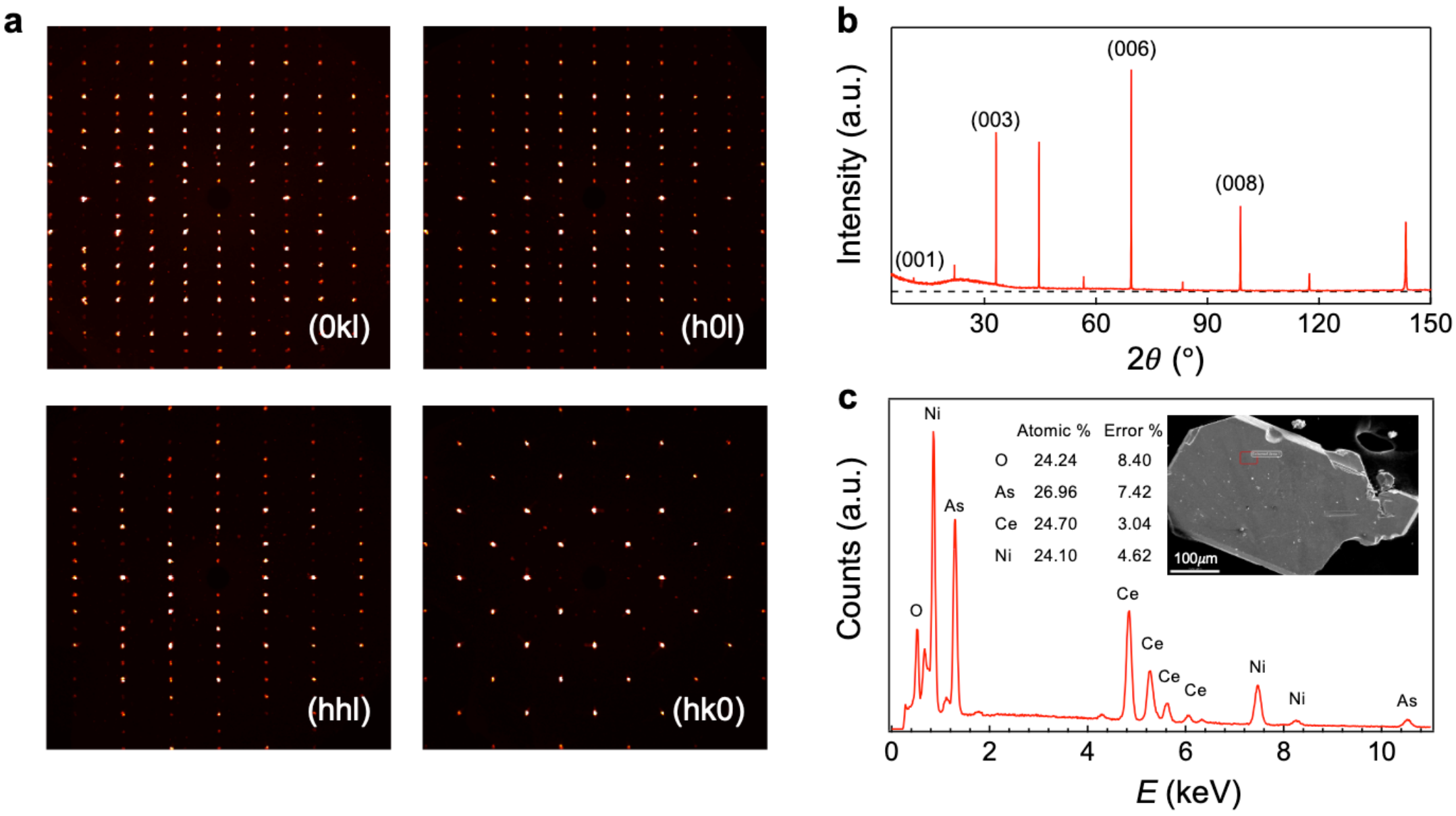

**Extended Data Fig. 1 | Structural and compositional characterization of CeNiAsO single crystals. a,** Single-crystal x-ray diffraction patterns measured on the (0kl), (k0l), (hhl), and (hk0) crystal planes at 300 K. **b,** X-ray diffraction result of the exposed *ab* plane. **c,** Energy-dispersive x-ray spectroscopy results, indicating an approximately 1:1:1:1 atomic ratio of Ce, Ni, As, and O.

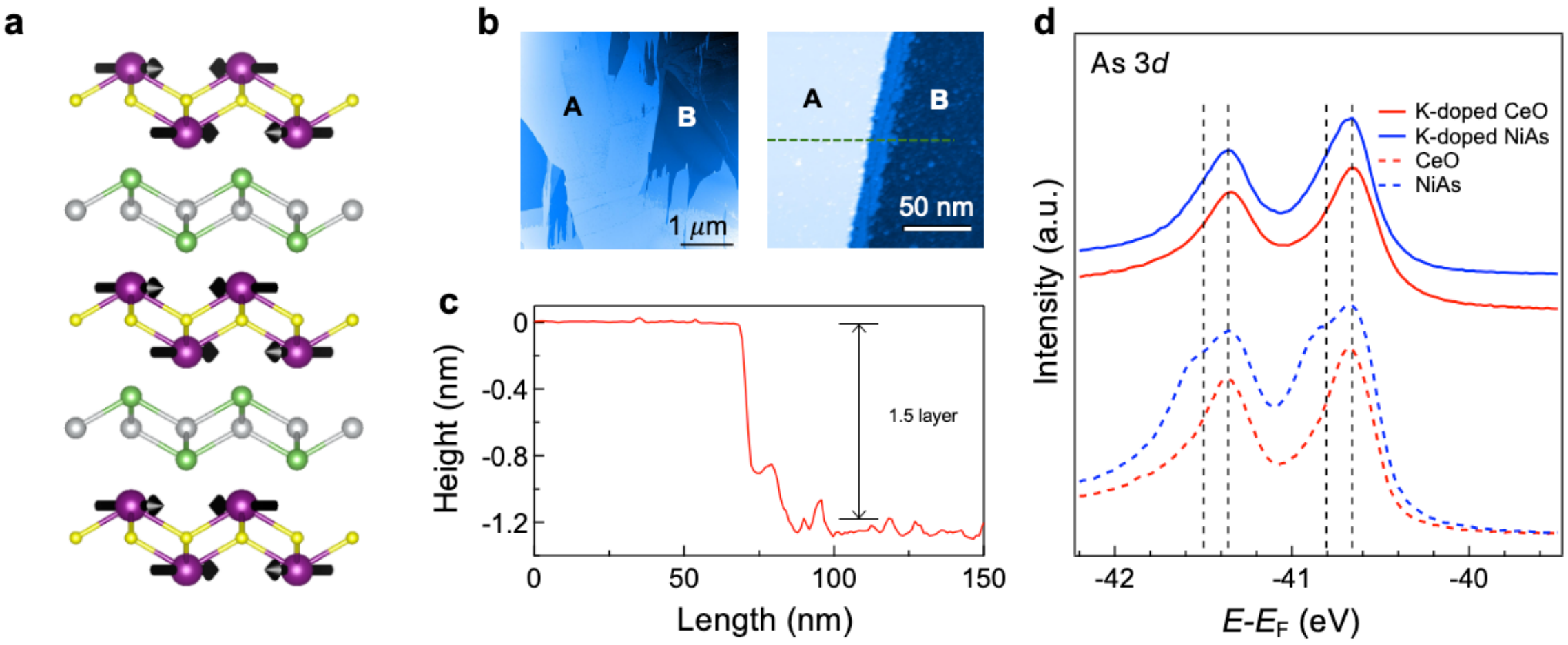

**Extended Data Fig. 2 | Two surface terminations of CeNiAsO single crystals. a,** Crystal structure of CeNiAsO, showing the two possible surface terminations: CeO and NiAs. **b,** Scanning tunnelling microscopy (STM) images of the two terminations. Surface A is relatively smooth, whereas surface B is comparatively rough. **c,** Height profile taken along the dashed line indicated in **b**. The height difference of approximately 1.5 layers indicates that surfaces A and B correspond to different terminations. **d,** Core-level spectra of the As 3d orbitals for the CeO- and NiAs-terminations. The dashed and solid lines represent the spectra before and after K doping, respectively.

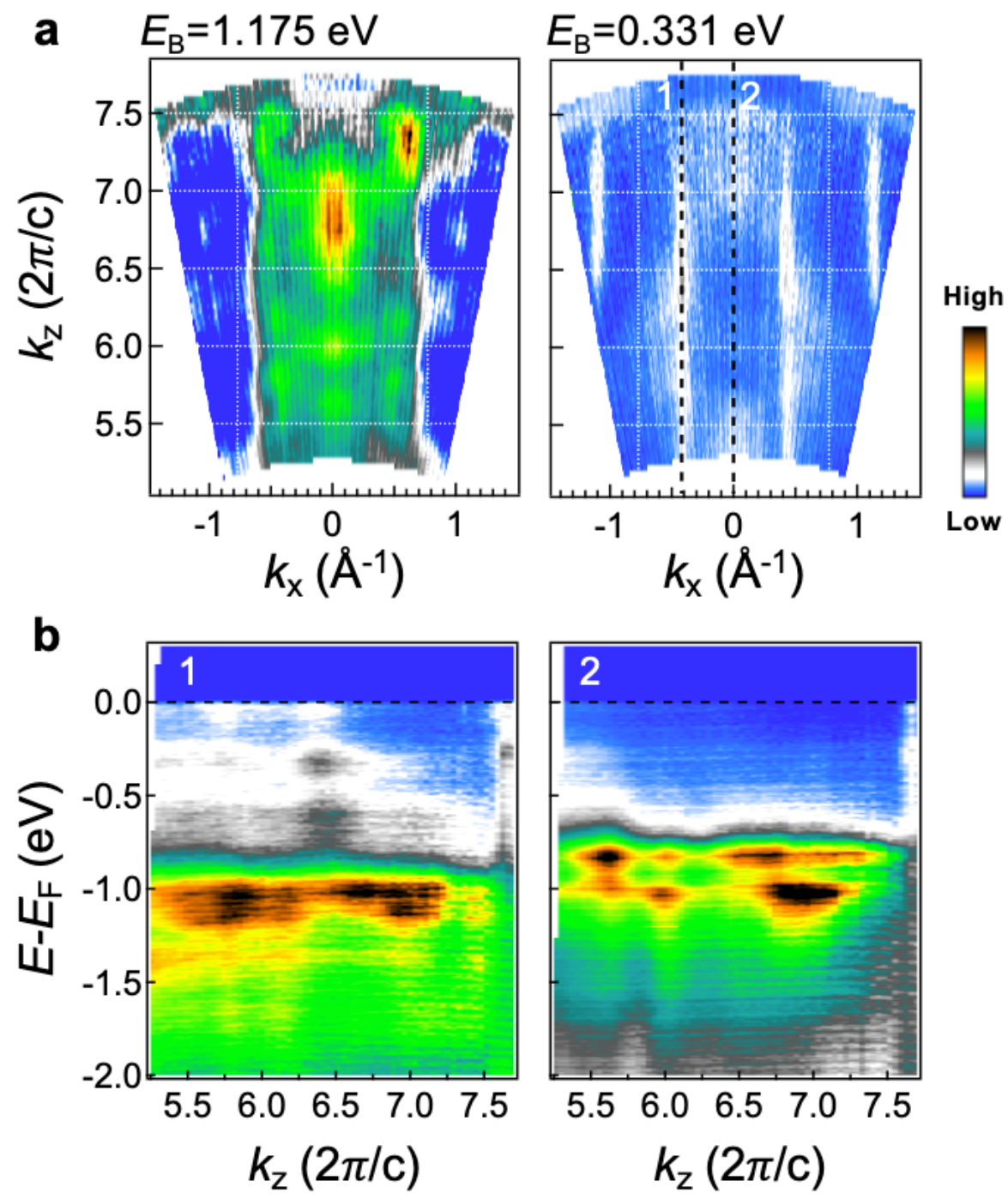

**Extended Data Fig. 3 | Electronic structure in the $k_x$-$k_z$ plane measured by photon energy-dependent ARPES.** **a,** Constant energy contours at binding energies of 1.175 eV and 0.331 eV. **b,** Band spectra corresponding to cuts 1 and 2 indicated in **a**, respectively. Inner potential of 19 eV is inferred.

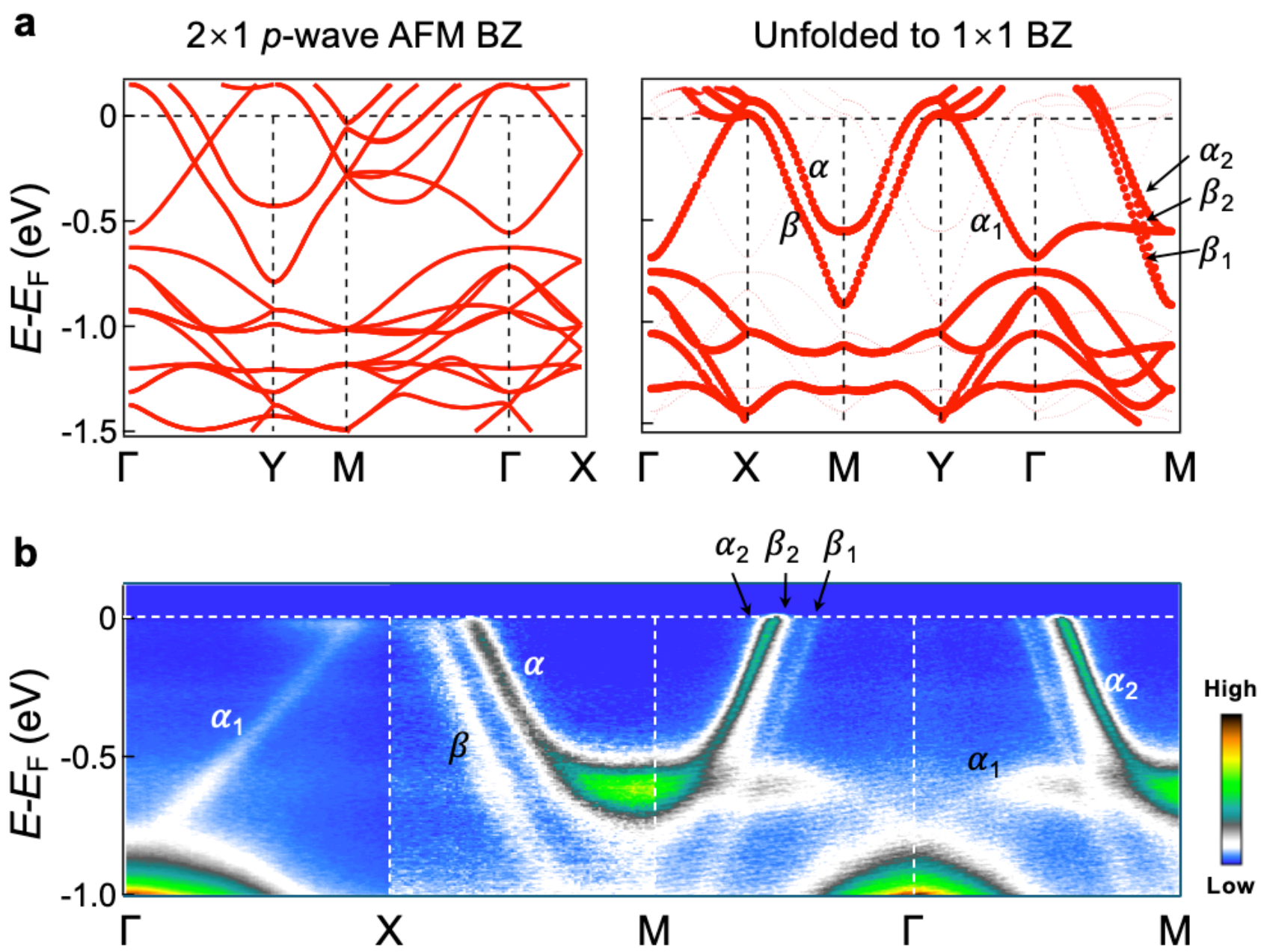

**Extended Data Fig. 4 | Electronic structure of CeNiAsO in the *p*-wave magnetic configuration. a,** Folded and unfolded calculated band structure, respectively. **b,** ARPES intensity plot along high symmetry path measured at 6.5 K.

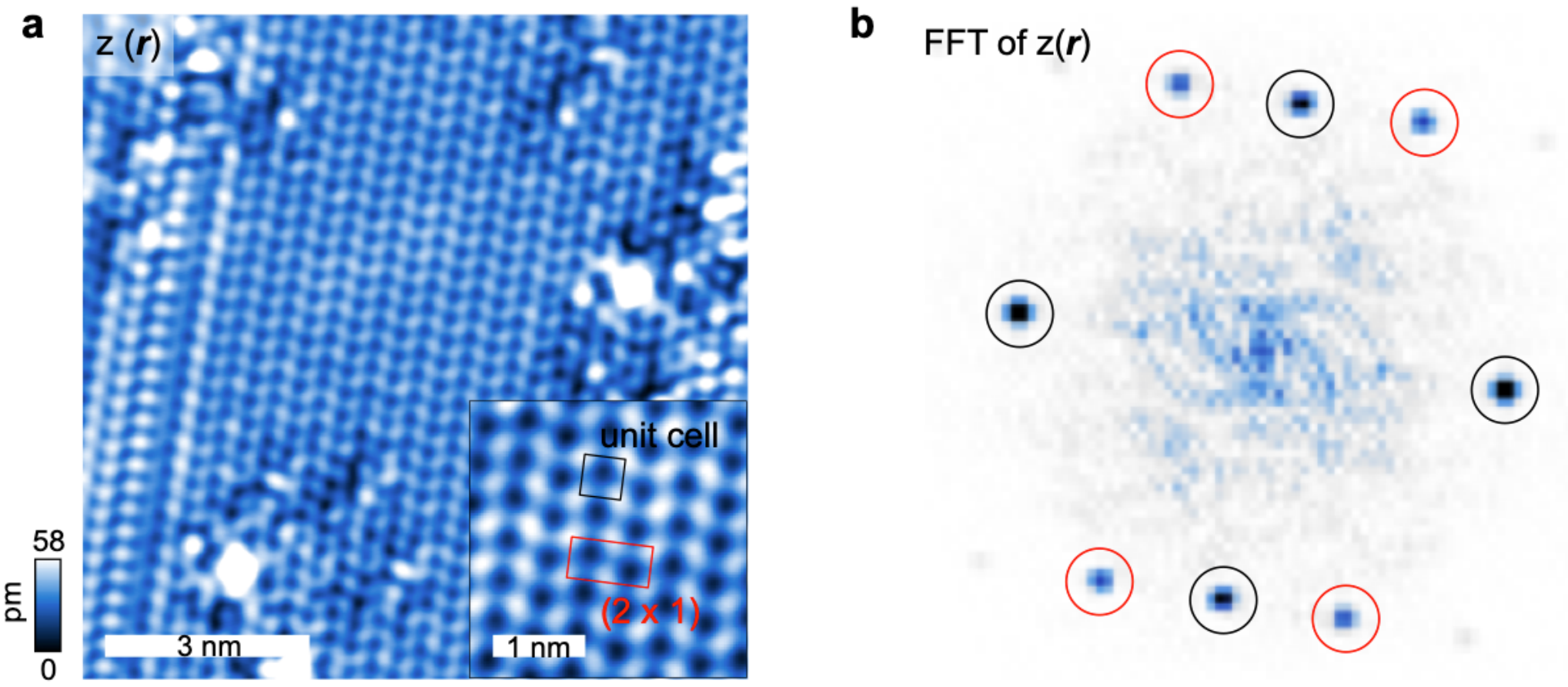

**Extended Data Fig. 5 | 2×1 reconstruction on surface A. a,** STM topographic image of the freshly cleaved surface A. The inset displays a zoomed-in area clearly revealing the 2×1 reconstruction. The black square and the red rectangle mark the unit cell and the reconstructed supercell, respectively. The topography is taken at $V_{bias}$ = -100 mV, $I_t$ = 100 pA and $T$ = 1.56 K. Scalebars: 3 nm and 1 nm (inset). **b,** Fast Fourier transform (FFT) image of topography in **a.** Vectors highlighted by black and red circles originate from the original lattice and reconstructed superlattice, respectively.

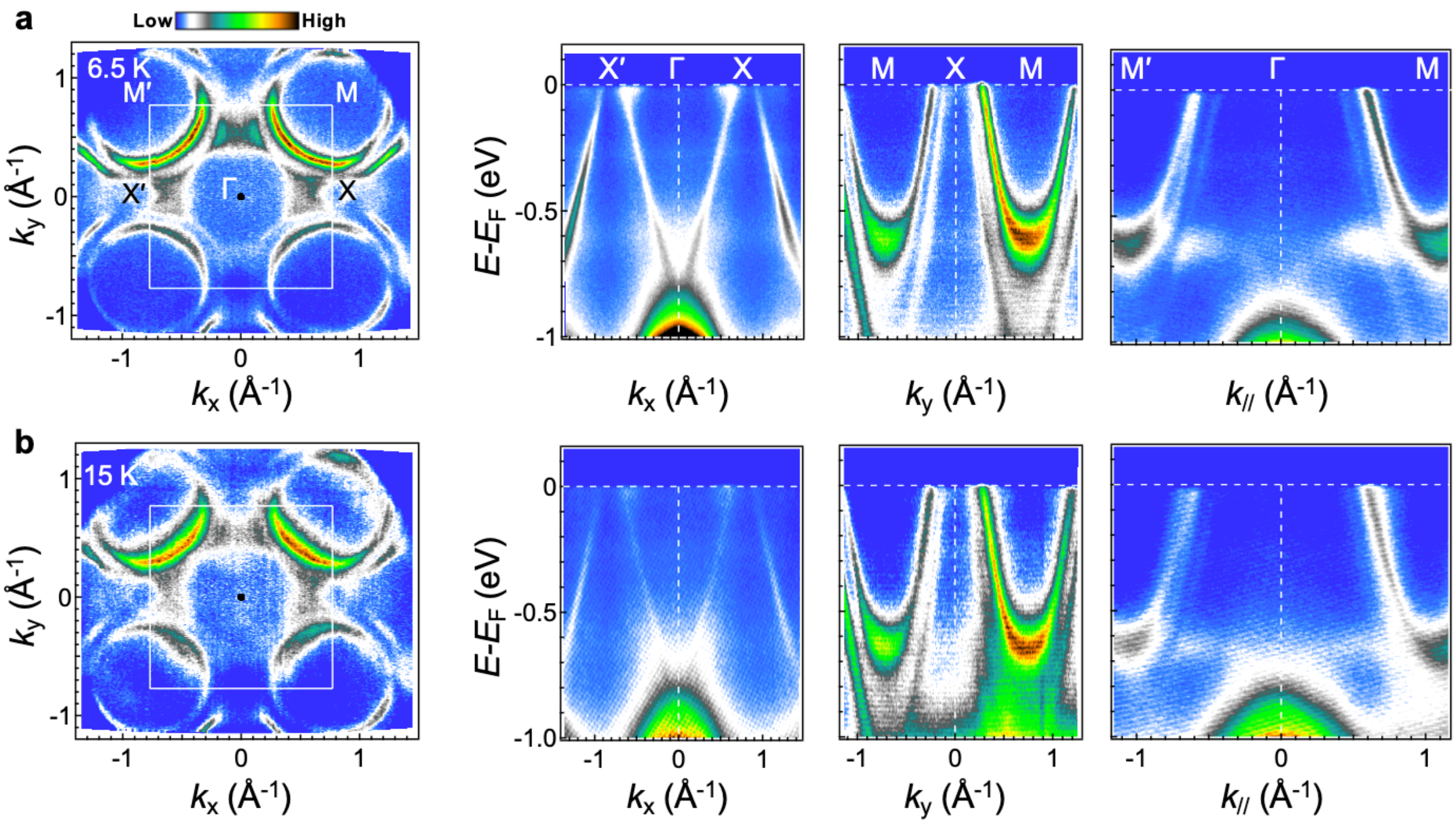

**Extended Data Fig. 6 | Temperature dependence of band structure. a-b,** Fermi surface and band spectra along high symmetry directions measured at 6.5 K and 15 K, respectively. The data taken in antiferromagnetic and paramagnetic states show essentially the same electronic structure.

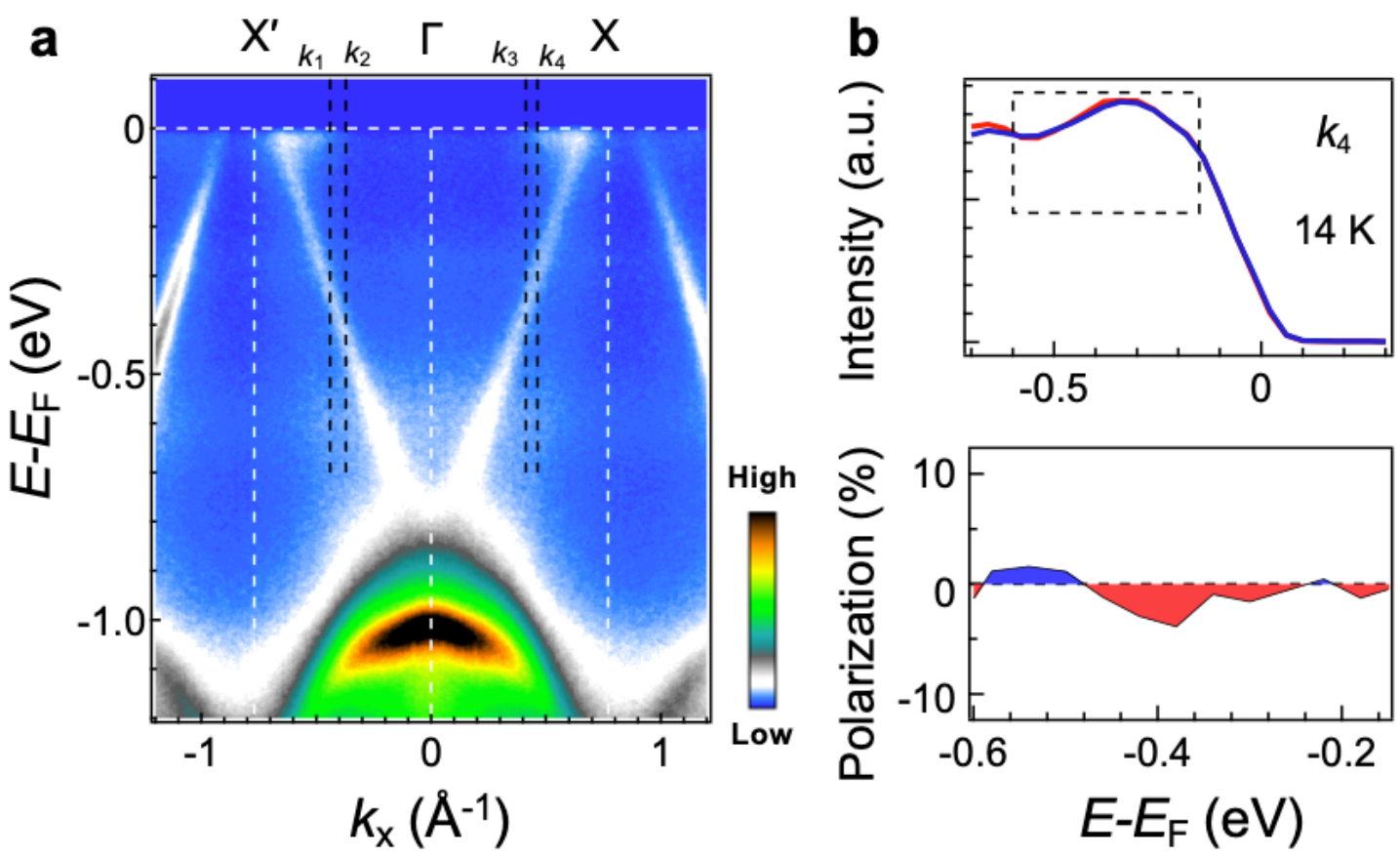

**Extended Data Fig. 7 | Spin-resolved ARPES results measured at 14 K. a,** ARPES spectra along X'-Γ-X. **b,** Spin-resolved energy-distribution curves and spin polarization corresponding to $k_4$ in **a**, showing spin polarization in the noise level.

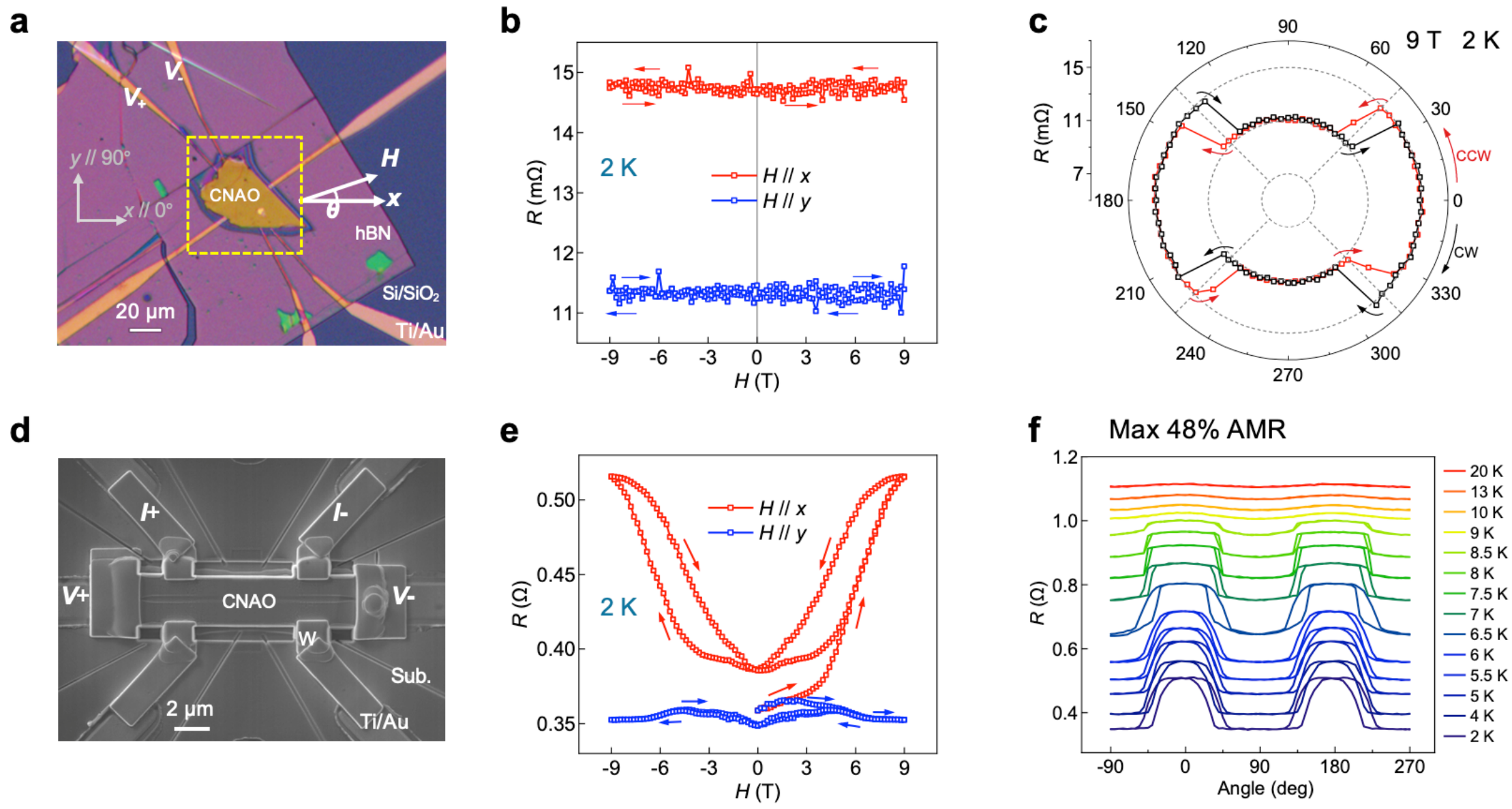

**Extended Data Fig. 8 | Anisotropic resistance in different devices. a,** Top view of the bulk device. The CeNiAsO flake is placed on the fabricated Ti/Au electrodes and hBN is employed as a capping layer to protect the sample. An in-plane magnetic field H is applied at an angle θ with respect to the current direction. **b,** *R*-*H* curves under magnetic field applied along *x*-axis (red) or *y*-axis (blue). The arrows indicate the field scan direction. **c,** Polar plot of the *θ*-dependent anisotropic resistance under a 9 T in-plane magnetic field at 2 K. The red (black) curve represents the magnetic field sweeping counterclockwise (clockwise). **d**. Scanning electron micrograph of the FIB Device B. **e**. *R*-*H* curves under magnetic field applied along *x*-axis (red) or *y*-axis (blue) at 2 K . The arrows indicate the field scan direction. **f**. The angle-dependent resistance under a 9 T in-plane magnetic field at different temperatures. The maximum magnetoresistance ratio calculated by $(R_{max}-R_{min})/R_{min}\times100\%$ is up to 48% at 2 K.